\documentstyle [aps,eqsecnum,twocolumn]{revtex}

\begin{document}

\title{Modified Extrapolation Length Renormalization Group Equation}
\author{Jacob Morris and Joseph Rudnick \\* {\em Department
of Physics, University of California, Los Angeles }}

\maketitle

\begin{abstract}

A modified renormalization group equation for the inverse extrapolation
length $c$ is derived by considering the phase shifts of order parameter
fluctuations.  The resulting non-linear equation is also derived using
standard methods and some additional assumptions.  The associated
renormalized flow $c(l)$ exhibits the correct behavior near both the
special and ordinary fixed points and in particular yields a canonical
scaling of $c$ with cross-over exponent $\phi_{\rm ord} = -\nu$ near the
ordinary transition.  \\*
\end{abstract}
Pacs: 05.70.Jk, 64.60.Ak, 64.60.Fr, 68.35.Rh

\section{Introduction}

Since the first renormalization group (RG) analysis of surface critical
behavior by Lubensky and Rubin\cite{lr1} a number of subsequent
advances\cite{dd1,bm,cg,dl} in the technique have enabled computation of
exponents to $O(\epsilon^{2})$, critical amplitudes, and various cross-over
functions\cite{dd1,jg,g,gm,dgs}.  In particular, Diehl and Dietrich have
systematically developed a formalism whereby the power and elegance of the
field theoretic method has been fully exploited.

For the case of an $O(1)$ system confined to the half-space $z>0$ it
suffices to consider the reduced Hamiltionian
\begin{equation}
{\cal H} [ \phi] = \int d^{d}x \left[ \frac{1}{2}(\nabla{\phi})^{2} +
\frac{r}{2}\phi^{2} + \frac{u}{4!}\phi^{4} \right] + \int dS \frac{c}{2}
\phi^{2}
\label{effH}
\end{equation}
the presence of the bounding surface manifesting itself as an additional
surface interaction\cite{dl}.  The parameter $c$ takes account of the local
enhancement of the reduced temperature in the vicinity of the surface.  At
lowest order the surface term results in the boundary condition $\phi'(0) =
c \phi (0)$ and thus $1/c$ corresponds to the distance over which the order
parameter falls to zero when extrapolated away from the surface.  For $c>0$
the surface orders with the bulk while for $c<0$ there is an enhanced
tendency to order at the surface.  The ``special" transition with $c=0$
divides these two regimes whereas the ``ordinary" transition with
$c=\infty$ corresponds to a state where ordering at the surface is
completely supressed.

An issue of considerable interest is the manner in which various quantities
behave close to the ordinary transition and to this end an expansion in the
{\it bare} extrapolation length $1/c$ has been developed\cite{dde,dd3}.
Among the results is the finding that at the ordinary point energy related
quantities involving $\phi^{2}$ averages exhibit behavior characterized by
relations involving {\it bulk} exponents\cite{be}.  This in turn is a
direct consequence of a vanishing anomalous exponent $\eta_{c}$ associated
with the extrapolation length.

The canonical scaling of $c(l)$ near the ordinary point is interesting in
light of the fact that all analyses addressing the cross-over behavior from
the special to ordinary point\cite{g,jg,gm} have utilized a {\it linear} RG
equation which in dimensional regularization reads
\begin{equation}
\frac{d c}{dl} = (1 + \eta_{c}) c \label{eq:lc}
\end{equation}
with $e^{l}$ corresponding to the block spin size.  For {\it finite} $c(l)$
Eq. (\ref{eq:lc}) results from a straightforward application of the field
theoretic method to a bulk system with a planar bounding
surface\cite{dd1,dl}.  Clearly equation (\ref{eq:lc}) yields a flow which
is independent of the proximity to the special transition and thus does not
display the expected classical behavior at large $c(l)$.  It is readily
verified, however, that this disparity in scaling behavior from that
inferred from the $1/c$ expansion is compensated for by the crossover
functions exhibiting logarithmic singularities at large $c(l)$.  When
exponentiated, these singularities lead to powers of $c(l)$ that in effect
undo the incorrect large $c$ behavior of Eq. (\ref{eq:lc}) and in turn lead
to the appropriate exponents at the ordinary point.

With the above in mind a question of immediate interest is to what extent
it is possible to deduce a flow for $c(l)$ that correctly interpolates
between the special and ordinary points.  Constructing such a flow is not
immediately obvious since for finite $c$ the linear RG Eq. (\ref{eq:lc})
results from the standard program of renormalizing all relevant surface
operators.  It has been pointed out\cite{dde}, however, that near the
ordinary point additional care must be used in categorizing operators as
relevant or irrelevant in the RG sense.  In particular, in the context of
the $1/c$ expansion it happens that insertions of the formally irrelevant
interaction $(\partial_{n} \phi)^{2}$ must be considered\cite{dde}.

In the following we outline an approach based on the physical notion
of an extrapolation length that gives rise to a non-linear RG equation
exhibiting the correct behavior both at the special and ordinary fixed
points.  As in the case of the $1/c$ expansion the operator
$(\partial_{n} \phi)^{2}$ is found to play a important role in the
analysis.  In addition to yielding an RG flow with the sought-after
behavior at both fixed points, the method further elucidates the
connection between the extrapolation length and parameter $c$.  This
insight is of interest in its own right since conventional wisdom
holds that the connection loses meaning beyond mean-field
theory\cite{dl}.  Making use of an analysis based on phase shifts, we
will demonstrate that there is a means of extending the notion of an
extrapolation length that remains intact when fluctuations are taken
into account.  This approach may have the potential for further
development since the use of phase shifts in many-body systems is a
concept often incorporated into perturbative analyses of arbitrary
order\cite{fw}.

\section{Scattering Formulation}

Our analysis begins with the reduced Hamiltonian of Eq.  (\ref{effH})
with the undestanding that all volume integrations are to be taken
over the half-space $z> 0$.  Within mean field theory, straightforward
variation of the Hamiltonian (\ref{effH}) gives rise to the boundary
condition
\begin{equation}
c \phi (0) = \phi'(0) \label{eq:bc}
\end{equation}
Beyond mean field theory there is an approach in which the connection
between $c$ and an extrapolation length is still apparent.  The key
observation is that the oscillatory nature of the modes leads to the
extrapolation length manifesting itself as a {\it phase shift}.  In
particular, taking the free Hamiltonian ${\cal H}_{0}$ to be
\begin{equation}
{\cal H}_{0} = \int d^{d}x \frac{1}{2}\left[ (\nabla{\phi})^{2} + r
\phi^{2} \right] + \int dS \frac{c}{2} \phi^{2}
\end{equation}
the modes which diagonalize ${\cal H}_{0}$ satisfy the boundary condition
Eq. (\ref{eq:bc}) and are of the form
\begin{equation}
\phi_{k} = e^{-ikz} - f_{k} e^{i kz}
\end{equation}
where the scattering amplitude $f_{k}$ and phase shifts are given by
\begin{equation}
f_{k} = \frac{c + ik}{c-ik} = e^{2i \delta_{k}}, \; \; \; \tan \delta_{k} =
k/c
\end{equation}
At this level of approximation it is apparent that the presence of the
surface interaction serves as an effective scattering potential
characterized by phase shifts $\delta_{k}$.  Conversely, given phase shifts
$\delta_{k}$ the associated inverse extrapolation length satisfies
\begin{equation}
\lim_{k \rightarrow 0} \frac{\delta_{k}}{k} = \frac{1}{c} \label{eq:c}
\end{equation}
The inclusion of fluctuations will alter the effective surface potential.
To ascertain how fluctuations influence the extrapolation length one can
appeal to the manner in which the phase shifts are modified and then use
Eq. (\ref{eq:c}).

We now carry out this program by addressing the lowest order, one-loop
corrections.  For a given self energy $\Sigma$ the modes satisfy
\begin{equation}
-\frac{d^{2} \phi}{d z^{2}} + (t + q^{2}- E_{k}) \phi = - \sigma (z) \phi
\label{eq:sc}
\end{equation}
where $\sigma (z) = \Sigma (z) - \Sigma (\infty)$, and $t = r + \Sigma
(\infty)$ is the suitably shifted bulk reduced temperature.  Since we
will ultimately be implementing the RG, let us assume that $\sigma
\sim \epsilon$ which then permits Eq.  (\ref{eq:sc}) to be solved
using standard perturbation theory.  In the present circumstance in
which results appropriate to one-loop order are sought, first order
perturbation theory suffices.  We are led to a modified scattering
amplitude
\begin{equation}
f_{r} = f - \frac{1}{2ik} \int dz [ \phi^{0}_{k} (z)]^{2} \sigma (z)
\end{equation}
The corresponding fluctuation corrected $c_{r}$ is obtained by considering
the small $k$ limit of Eq. (\ref{eq:c}).  Noting that $f \simeq 1 + 2ik/c$
it follows that
\begin{equation}
\frac{1}{c_{r}} = \frac{1}{c} - \int dz (z + 1/c)^{2} \sigma (z)
\label{eq:ci}
\end{equation}
which is valid to $O(\epsilon)$.  In the event that higher order
corrections are desired it is necessary to take account of additional
perturbative corrections to Eq. (\ref{eq:sc}).

\section{Renormalization Group}

The above results can now be used to determine the renormalization
group equation for $c(l)$.  We proceed with a momentum-shell approach.
To this end, we note that translational invariance in directions
parallel to the surface allows one to write
\begin{equation}
\phi ({\bf x}) = \sum_{q} \phi_{q} (z) e^{i {\bf q} \cdot {\bf y}}
\end{equation}
Integrating out all modes with parallel momentum in the shell $e^{-\Delta
l}<q < 1 $ and using the result that the averages obey
\begin{equation}
\langle \phi_{q} (z) \phi_{-q} (z') \rangle = \frac{1}{2 \kappa} \left[
e^{-\kappa |z-z'|} - a \, e^{-\kappa(z+z')} \right]
\end{equation}
with
\begin{equation}
a = \frac{c - \kappa}{c + \kappa}
\end{equation}
and $\kappa^{2} = q^{2} + t$, one finds for the subtracted self energy
\begin{equation}
\sigma (z) = - \frac{u K_{d-1}}{4 \kappa_{1}} \Delta l \;
 a_{1} e^{-2 \kappa_{1} z} \label{eq:sg}
\end{equation}
where the 1-subscript refers to all quantities being evaluated at $q = 1$.
Rescaling lengths so that $ c \rightarrow e^{\Delta l} c$, one arrives at
the RG equation
\begin{equation}
\frac{dc}{dl} = c - \frac{u^{*} K_{d-1}}{8} \left[\frac{c
-\kappa_{1}}{\kappa_{1}^{3}} + \frac{c^{2}}{2\kappa_{1}^{4}} \frac{(c -
\kappa_{1})}{(c + \kappa_{1})}
\right] \label{eq:rc}
\end{equation}
where $u$ has now been set to its fixed point value $u^{*} K_{d-1} = 8
\epsilon/3$.  For comparison we note that the corresponding equation
resulting from a standard momentum shell approach\cite{jj} in which only
the interactions $\phi^{2}, \phi \partial_{n} \phi$ are considered reads
\begin{equation}
\frac{dc}{dl} = c - \frac{u^{*} K_{d-1}}{8} \left[\frac{c
-\kappa_{1}}{\kappa_{1}^{3}} \right] \label{eq:rc1}
\end{equation}
Equation (ref{eq:rc1}) is the hard cut-off version of the
dimensionally regularized result (\ref{eq:lc}).  This latter equation
being {\it linear} in $c$ implies a flow
\begin{equation}
c(l) = e^{l \phi/ \nu}
\end{equation}
that is independent of the proximity to special transition.
Deviations between equations (\ref{eq:rc1}, \ref{eq:rc}) begin to
appear when $c(l) \sim 1$.  The third non-linear term appearing in Eq.
(\ref{eq:rc}) for finite $c$ is ultraviolet convergent and corresponds
to the inclusion of contributions from the formally irrelevant
$(\partial_{n} \phi)^{2}$ vertex.  However, if this last term is
expanded in $1/c$, it is clear that corrections to the shift and
exponent of $c$ occur, and that successive terms become increasingly
ultraviolet divergent.

Equation (\ref{eq:rc}) leads to some interesting results, which we now
address.  For the sake of illustration assume that the system is close
enough to criticality so that cross-over to the ordinary point has
already occurred while $r(l) \ll 1$.  In this case Eq.  (\ref{eq:rc})
reduces to
\begin{equation}
 \frac{dc}{dl} = c - \frac{u^{*} K_{d-1}}{8} \left[c -1 + \frac{c^{2}}{2}
 \frac{(c - 1)}{(c + 1)}
\right] \label{eq:rc0}
\end{equation}
Although it is possible to solve Eq.  (\ref{eq:rc0}) exactly, only
results accurate to $O(\epsilon)$ will be considered.  There are
several ways to go about solving Eq.  (\ref{eq:rc0}), one of which
proceeds by iteratively solving the differential equation to
$O(\epsilon)$.  This leads to the explicit solution
\begin{eqnarray}
c(l) & = & b(l) - \frac{u^{*} K_{d-1}}{8}
\left[1 + \frac{b(l)^{2}}{2}
 - b(l) \ln (1 + b(l)) \right].  \\ \label{eq:cren} b(l) & = & b(0)
 e^{(1+\eta_{c})l} \nonumber \\ \nonumber
\end{eqnarray}
with $ b(0) = c(0) + u^{*} K_{d-1}/8$, and $\eta_{c} = -\epsilon/3 $.
Another method approximates the roots to the resulting cubic on the right
hand side of Eq. (\ref{eq:rc0}) and leads directly to the implicit form
\begin{equation}
\left[\frac{c(l)- \eta_{c}}{c(0) - \eta_{c}} \right]^{1- \eta_{c}}
\left[\frac{1+ c(l) + \eta_{c} }{1+c(0) + \eta_{c}}
\right]^{\eta_{\rm c}} \left[\frac{2/\eta_{c} + c(0)}{2/\eta_{c} + c(l)}
\right] = e^{l} \label{eq:creni}
\end{equation}
which can also be shown to follow from exponentiation of (\ref{eq:cren}).
Inspection of the above results reveals that for $c(l) \ll 1$ the flow is
characterized by $\eta_{c} \neq 0$ while close to the ordinary point $c(l)
\sim e^{l}$ thus implying a vanishing $\eta_{c}$.  Stated differently, the
cross-over exponent for the extrapolation length $\lambda (l) = 1/c(l)$ at
the ordinary point is $\phi_{ord} = - \nu$.  Another interesting feature of
Eq. (\ref{eq:creni}) is that it yields an ordinary fixed point of order
$c(\infty) \sim 1/\epsilon $.

Past analyses, employing the momentum-shell technique to surface
related phenomena, have encountered various technical
difficulties\cite{lr1,cg}.  We, therefore, first consider how this
method leads to the standard linear equation (\ref{eq:lc}) before
attempting an alternate derivation of the modified RG equation
(\ref{eq:rc}).  As degrees of freedom are integrated out, additional
interactions are generated.  This is accommodated by taking the
surface interaction to be of the form
\begin{equation}
V(z) = \sum_{m} v_{m} \delta^{(m)}(z) \label{eq:delt}
\end{equation}
with $\delta^{(m)}(z)$ referring to a $m^{\rm th}$ derivative.  For given
$V(z)$ the coefficients $v_{m}$ are determined by
\begin{equation}
v_{m} = \frac{(-)^{m}}{m!} \int_{0}^{\infty} z^{m} V(z) dz \label{eq:mom}
\end{equation}
Consider the one-loop contribution, which results in the surface
interaction $V(z) = \sigma (z)$ given by Eq.  (\ref{eq:sg}).  After
rescaling the surface spins by a factor $e^{\Delta l(1-\eta_{1})/2 }$
one finds that the coefficients $v_{m}$ satisfy the recursion
relations:
\begin{equation}
\frac{d v_{m}}{dl} = (1-m-\eta_{1}) v_{m} - \frac{u K_{d-1}}{2}
\frac{(-)^{m}
a_{1}}{(2 \kappa_{1})^{m+2}} \label{eq:vm}
\end{equation}
The vertex involving $ \phi^{2} \delta'(z)$, or equivalently
$\delta(z) \phi \partial_{n} \phi$ results from the boundary term
associated with $ (\nabla \phi)^{2}$.  Analogous to what is done in
bulk phenomena the factor $\eta_{1}$ is chosen so that $v_{1} = 1/2$
remains fixed.  This leads to the result
\begin{equation}
\eta_{1} = \frac{u^{*} K_{d-1}}{8 \kappa_{1}^{3}} a_{1}
\label{eq:et}
\end{equation}
When this value for $\eta_{1}$ is inserted into Eq.  (\ref{eq:vm}) for
$v_{0}$, one ends up with the linear RG equation (\ref{eq:lc}).  It is
interesting that the non-linearity associated with the factor $a_{1}$
is entirely cancelled.  Inspection of Eq.  (\ref{eq:vm}) governing
$v_{m}$ reveals that all interactions with $m \ge 2$ are irrelevant.
However, in the context of calculating various scaling functions, these
interactions with $m \geq 2$ must, in fact, be considered to account for
all $O(\epsilon)$ contributions\cite{jj}.

It is possible under certain circumstances to interpret the
contribution to $c$ from $\eta_{1}$ as feeding in from the $v_{1}$
vertex.  This becomes evident upon considering the contribution each
surface term makes when inserted into a propagator with {\it legs off
the surface}.  Recall that the $m^{\rm th}$ vertex involves a factor
$\delta^{(m)}(z)$, which leads, after an integration by parts, to an
interaction $\delta (z) \partial_{z}^{m} [\phi^{2}]$.  The boundary
condition then effectively relates this to a term proportional to
$\delta (z) \phi^{2}$ and leads to a contribution from $v_{m}$ feeding
into the recursion for $c$.  In such case it is possible that the
higher order surface interactions will influence the behavior of $c$.

To see how the above reasoning leads to the modified RG Eq. (\ref{eq:rc})
assume that all two point interactions ultimately will modify a propagator
with legs off the surface and that these legs each carry a transverse
momentum {\it q} with associated factor $\kappa_{0} = \sqrt{q^{2} + r(l)}$.
The assumption that the legs are off the surface effectively leads to the
replacements:
\begin{eqnarray}
 \partial_{n}^{2 m +1} \phi & \rightarrow & c \kappa_{0}^{2m}\phi \\
\partial_{n}^{2m} \phi & \rightarrow & \kappa_{0}^{2m} \phi \label{eq:cp}
\end{eqnarray}
in all surface interactions.  Note that the $\delta$ function
singularity associated with two or more derivatives makes no
contribution because of the fact that the legs are off the surface.
When momenta in a thin shell are integrated out, each $v_{m}$ receives
a contribution
\begin{equation}
\Delta v_{m} = \frac{\Delta l}{2} u K_{d-1} \frac{(-)^{m} a_{1}}{(2
\kappa_{1})^{m+2}}\label{eq:v}
\end{equation}

For the moment, we will ignore any contribution from an anomalous
surface spin rescaling factor $\eta_{1}$.  Integrating by parts and
invoking the correspondence (\ref{eq:cp}), the $v_{1}$ interaction
leads to a term $ \Delta v_{1} \delta (z) c \phi^{2}$.  Similarly, the
$v_{2}$ vertex involves $\partial^{2} [\phi^{2}]$ and thus leads to a
term $2 \Delta v_{2} \delta (z) (c^{2} + \kappa_{0}^{2}) \phi^{2}$.
It follows that the total effective contribution from the $v_{1},
v_{2}$ interactions to $v_{0}$ is
\begin{equation}
\Delta v_{0} = \Delta v_{1} c + 2 \Delta v_{2} (c^{2} + \kappa_{0}^{2})
\label{eq:dv}
\end{equation}

Rescaling spins and lengths, using Eq.  (\ref{eq:v}), and for the
moment ignoring the last term involving $\kappa_{0}$, one arrives at
the modified RG Eq.  (\ref{eq:rc}).  Alternatively, identifying
$\Delta v_{m}$ with the moments of $V(z)$ using Eq.  (\ref{eq:mom})
one recovers Eq.  (\ref{eq:ci}) derived from the scattering theory
approach.  Though $\eta_{1}$ was ignored, the final result is the same
when anomalous spin rescaling is included.  If spins are rescaled so
that $v_{1}$ remains fixed, while there is no contribution to $v_{0}$
in the form of $\Delta v_{1}$, there is a contribution from $\eta_{1}$
which, because of Eq.  (\ref{eq:et}), yields an identical result.

The above analysis arbitrarily neglected the contributions from the $v_{2}$
vertex in addition to all interactions with $m >2$.  To determine under
what circumstances this is justified note that the vertex $v_{m}$ makes a
contribution to $\Delta v_{0}$ of order

\begin{equation}
\frac{\kappa^{m-2}_{0}}{\kappa_{1}^{m +2}} \left[(c + \kappa_{0}
)^{2} + (-)^{m} (c-\kappa_{0})^{2} \right] a_{1}
\end{equation}
and becomes increasingly negligible for $\kappa_{0} \ll \kappa_{1}
\sim 1$.  This latter condition is satisfied sufficiently close to the
critical point when leg momenta ${\mbox q \ll 1}$.  For the current
situation of interest here this condition is well satisfied.  However,
in view of this assumption, our derivation strictly applies only to
the RG Equation (\ref{eq:rc0}), in which $r(l)$ was neglected.

Generally, when $r(l)$ is not negligible it is possible to sum the
higher order corrections that were neglected in the above analysis.
However, the resulting equation differs from that found using phase
shifts(\ref{eq:rc}).  The differences arise from the two methods
reflecting different conventions on the finite part of $c(l)$.  This
is, of course, compensated for by making a correspondingly different
subtraction, depending on which flow is used.

\section{Concluding Remarks}

We have presented a method for identifying an effective surface enhancement
$c(l)$ which utilizes the scattering phase shifts of the localized part of
the self energy $\sigma (z)$.  The resulting cross-over behavior in $c(l)$
is found to arise from the inclusion of contributions of various higher
order surface interactions, in particular $(\partial_{n} \phi)^{2}$.  It is
interesting that the relatively simple connection involving phase shifts
implicitly includes such higher order corrections.  Furthermore, the method
appeals to characteristics of the entire (smooth) surface interaction
rather that its constituent localized (delta function) pieces and thus may
lead to further insights into surface phenomena.  Indeed, though often
convenient, the use of hyper-localized surface distributions is somewhat
unphysical and occasionally leads to pathological quantities requiring
special limiting procedures and interpretations.

The task of determining a scaling field with the correct scaling
behavior at both fixed points is important in its own right.  We have
performed a preliminary analysis of the scaling functions for the
surface susceptibility and surface free energy and find, as expected,
that the use of a modified flow similar to Eq.  (\ref{eq:cren})
eliminates the logarithmic singularities otherwise found in these
quantities\cite{jg,gm}.  This in turn suggests that the logarithmic
singularities in these two quantities are due entirely to the
cross-over in $c(l)$.

Within his calculation of the local susceptibility Goldschmidt\cite{g} also
addressed the exponentiation of the logarithmic singularities appearing in
this quantity.  In this particular scaling function, however, the
introduction of our modified flow is not sufficient to eliminate the
singularity.  We have also verified this is also the case for the layer
susceptibility\cite{jj}.  This is to be expected, however, since both these
quantities involve at least one external point on the surface.

\end{document}